\documentclass[useAMS,usenatbib,usegraphicx]{mn2e}

\usepackage{graphicx} 
\usepackage{amsmath} 
\usepackage{amssymb}
\usepackage{color}

\newcommand{\rtwo}{\rm{r}_{200}} 
\newcommand{\Mtwo}{\rm{M}_{200}}
\newcommand{\vtwo}{\rm{v}_{200}}
\newcommand{\stwo}{\sigma_{200}}
\newcommand{\slos}{\sigma_{\rm{los}}}
\newcommand{\soneD}{\sigma_{\rm{1D}}} 
\newcommand{\Msun}{\rm{M}_\odot}
\newcommand{\vel}{\,{\rm km\,s^{-1}}}

\newcommand{\mincir}{\raise
  -2.truept\hbox{\rlap{\hbox{$\sim$}}\raise5.truept \hbox{$<$}\ }}
\newcommand{\magcir}{\raise
  -2.truept\hbox{\rlap{\hbox{$\sim$}}\raise5.truept \hbox{$>$}\ }}

\title[Velocity dispersion -- mass scaling relation]{The relation
  between velocity dispersion and mass in simulated clusters of
  galaxies: dependence on the tracer and the baryonic physics}

\author[E. Munari et al.]
       {E. Munari$^{1,2}$,
       A. Biviano$^2$,
       S. Borgani$^{1,2,3}$,  G. Murante$^2$ \& D. Fabjan$^4$\\
$^1$Astronomy Unit, Department of Physics, University of Trieste,
  via Tiepolo 11, I-34131 Trieste, Italy (munari@oats.inaf.it)\\ 
$^2$INAF/Osservatorio Astronomico di Trieste, via Tiepolo 11,
  I-34131 Trieste, Italy  (biviano, borgani, murante@oats.inaf.it) \\
$^3$INFN – Istituto Nazionale di Fisica Nucleare, Trieste, Italy\\
$^4$ Center of Excellence SPACE-SI, A\v{s}ker\v{c}eva 12, 
1000 Ljubljana, Slovenia (dunja.fabjan@space.si)\\
 }

\begin{document}

\date{Accepted 2013 January 8. Received 2012 December 14; in original form 2012 April 23}

\pagerange{\pageref{firstpage}--\pageref{lastpage}} \pubyear{2012}

\maketitle

\label{firstpage}

\begin{abstract}
We present an analysis of the relation between the masses of cluster-
and group-sized halos, extracted from $\Lambda$CDM cosmological N-body
and hydrodynamic simulations, and their velocity dispersions, at
different redshifts from $z=2$ to $z=0$.  The main aim of this
analysis is to understand how the implementation of baryonic physics
in simulations affects such relation, i.e. to what extent the use of
the velocity dispersion as a proxy for cluster mass determination is
hampered by the imperfect knowledge of the baryonic physics.  In our
analysis we use several sets of simulations with different physics
implemented: one DM-only simulation, one simulation with non-radiative
gas, and two radiative simulations, one of which with feedback from
Active Galactic Nuclei.  Velocity dispersions are determined using
three different tracers, dark matter (DM hereafter) particles,
subhalos, and galaxies.

We confirm that DM particles trace a relation that is fully consistent
with the theoretical expectations based on the virial theorem,
$\sigma_v \propto M^\alpha$ with $\alpha = 1/3$, and with previous
results presented in the literature.  On the other hand, subhalos and
galaxies trace steeper relations, with velocity dispersion scaling
with mass with $\alpha>1/3$, and with larger values of the
normalization. Such relations imply that galaxies and subhalos have a
$\sim10$ per cent velocity bias relative to the DM particles, which
can be either positive or negative, depending on halo mass, redshift
and physics implemented in the simulation.

We explain these differences as due to dynamical processes, namely
dynamical friction and tidal disruption, acting on substructures and
galaxies, but not on DM particles. These processes appear to be more or
less effective, depending on the halo masses and the importance of
baryon cooling, and may create a non-trivial dependence of the
velocity bias and the $\soneD$--$\Mtwo$ relation on the tracer, the
halo mass and its redshift.

These results are relevant in view of the application of velocity
dispersion as a proxy for cluster masses in ongoing and future large
redshift surveys.

\end{abstract}

\begin{keywords}
galaxies: clusters: general --
galaxies: groups: general --
galaxies: kinematics and dynamics --
galaxies: evolution --
methods: numerical --
cosmology: theory
\end{keywords}

\section{Introduction}
\label{sect:intro}
Galaxy clusters provide a powerful means of tracing the growth of
cosmic structures and, ultimately, constraining cosmological
parameters \citep[e.g.][]{allen_etal11}.  A crucial aspect in the
cosmological application of galaxy clusters concerns the reliability
of mass estimates. Mass is not a directly observable quantity but can
be determined in several ways, e.g. by assuming the condition of
equilibrium of the intracluster plasma \citep[e.g.][]{EDGM02} or
galaxies \citep[e.g.][]{katgert04} within the cluster potential well,
or by measuring the gravitational lensing distortion of the images of
background galaxies by the cluster gravitational field
\citep[e.g.][]{Hoekstra03}.

These methods of mass measurement can only be applied to clusters for
which high quality data are available. When these are not available,
it is still possible to infer cluster mass from other observed
quantities, the so-called mass proxies, which are at the same time
relatively easy to measure and characterised by tight scaling
relations with cluster mass \citep[e.g.][]{kravtsov_borgani12}.
Examples of such mass proxies are the total thermal content of the
intra--cluster plasma, measured from either X--ray
\citep[e.g.,][]{kravtsov_etal06,stanek_etal10,fabjan_etal10} or
Sunyaev-Zel'dovich
\citep[e.g.][]{Arnaud+10,williamson_etal11,kay_etal11} observations,
the optical luminosity or richness traced by the cluster galaxy
population \citep[e.g.][]{PBBR07,rozo2009}, and velocity dispersion of
member galaxies \citep[e.g.][]{biviano06,saro2012}.

The use of velocity dispersion as a proxy for cluster mass is
particularly interesting in view of ongoing \citep[BOSS, see,
  e.g.][]{White+11} and forthcoming \citep[Euclid, see][]{Laureijs+11}
large spectroscopic galaxy surveys. It is crucial to understand
whether a cluster velocity dispersion measured on its member galaxies
is a reliable proxy for its mass.  Calibration of such scaling
relation can be based on detailed multi-wavelength observations of
control samples of galaxy clusters. On the other hand, detailed
cosmological simulations are quite useful to calibrate such scaling
relations independently from possible observational systematic effects
\citep[e.g.][and references therein]{borgani_kravtsov11}.

The implementation of baryonic physics can play a fundamental role in
these analysis.  In principle, since galaxies are nearly collisionless
tracers of the gravitational potential, one expects velocity
dispersion to be more robust than X-ray and SZ mass proxies against
the effects induced by the presence of baryons and by their thermal
history.

Using a set of cluster-sized halos extracted from a $\Lambda$CDM
cosmological simulation, \citet{biviano06} analysed the reliability of
the velocity dispersion as a mass proxy. They considered both DM
particles and simulated galaxies as tracers of the gravitational
potential of their host halo. They found that in typical observational
situations, the use of the line-of-sight velocity dispersion, $\slos$,
allows a more precise cluster mass estimation than the use of the
virial theorem. They used only one kind of simulation without
exploring different baryonic physics.

\cite{evrard08} analysed the $\stwo-\Mtwo$ \footnote{In the following
  we indicate with $\rm{r}_{200}$ the radius of the sphere drawn from
  the halo centre, which encloses a mean density of
  $200\times\rho_{\rm{c}}\rm{(z)}$, with $\rho_{\rm{c}}\rm{(z)}$ being
  the critical cosmic density at redshift $z$. The quantities with the
  subscript 200 have to be considered evaluated at, or within,
  $\rm{r}_{200}$. } relation using several cosmological simulations,
and showed that it is close to the virial scaling relation $\stwo
\propto \Mtwo^{1/3}$ across a broad range of halo masses, redshifts,
and cosmological models. When looking at simulated galaxies, some
studies found that they show a significant, albeit small, velocity
bias with respect to DM particles
\citep{diemand04,FKNG05,faltenbacher2006,faltenbacher2007,lau2010}.
These studies agreed that the amplitude of the velocity bias, i.e.
the ratio between the velocity dispersions of simulated galaxies and
DM particles, is not larger than $\approx 10$\%, but disagreed on
whether galaxies are positively (velocity bias $>1$) or negatively
(velocity bias $<1$) biased. The disagreement is unlikely to come from
resolution issues \citep{evrard08}. Other effects are more important
in affecting the value of the bias, such as the distance from the
cluster centre \citep{diemand04,gill04}, baryon dissipation and
redshift dependence \citep{lau2010}.

The way simulated galaxies are selected also has an important effect
on the amount of velocity bias.  By selecting simulated galaxies in
stellar mass, rather than in total mass, the velocity bias is strongly
reduced or even suppressed \citep{faltenbacher2006,lau2010}.  A
similar effect is seen when the selection of simulated galaxies is
based on their total mass at the moment of infall, which is found to
be proportional to the stellar mass
\citep{faltenbacher2006,lau2010,wetzel2010}. The proportionality
between total and stellar mass of a galaxy is lost after the galaxy
enters the cluster, because the DM halo is more easily stripped than
the stellar component by the cluster tidal forces
\citep{diemand04,boylan08,lau2010,wetzel2010}. Observational evidence
for tidal stripping of cluster galaxies has been obtained from lensing
studies \citep{NKS02,Limousin+07}. Tidal stripping is more effective
for galaxies moving at lower velocities \citep{diemand04}. When the
mass removed from a simulated galaxy by tidal stripping is such that
the galaxy mass drops below the resolution limit, the galaxy is
effectively tidally disrupted.  As the galaxies that are disrupted are
preferentially those of smaller velocities, the survivers will display
on average a larger velocity dispersion than DM particles, i.e. a
positive velocity bias \citep{faltenbacher2006}.

Another important process is dynamical friction
\citep{chandrasekhar1943,EF07}, which removes energy from a galaxy
orbit, bringing it closer to the cluster centre, and slowing down its
velocity \citep[e.g.][]{boylan08,wetzel2010}. If a sufficient number
of galaxies is slowed down by dynamical friction and survive both
tidal disruption and merging with the central galaxy, dynamical
friction might cause a negative velocity bias in the cluster galaxy
population.

All the processes discussed so far alter the dynamics of tracers like
galaxies, providing a source of uncertainty in the aforementioned
relation linking mass and velocity dispersion. With that comes the
need of further investigations about this topic, also because of the
different results found in the literature on the velocity bias of
cluster galaxies. The aim of this paper is to characterise the
relation between the velocity dispersion and the mass of simulated
halos spanning a wide mass range, from $\sim 10^{13} \Msun$ to $\ga
10^{15} \Msun$, at different redshifts (from $z=2$ to $z=0$), and
using different tracers, DM particles, subhalos, and galaxies, in
order to understand how reliable is the velocity dispersion as a proxy
for cluster masses. Simulations with different physics implemented are
used, in order to understand how different physical processes affect
the structures and hence the dynamics of tracers.

In this paper we do not consider observational biases, such as
projection effects and presence of interlopers
\citep{biviano06,saro2012}, but we focus on the effects due to the
physics and the implementation of baryonic physics in the
simulations. We find that such implementation affects the kinematics
of the systems. The analysis of observational effects must therefore
be based on simulations where baryonic physics is taken into account.

This paper is structured as follows. In Sect.~\ref{sect:simhalos} we
describe the simulations used for this work and define the samples
used in our analyses.  In Sect.~\ref{sect:svmrel} we determine the
relation between mass and velocity dispersion, and how it depends on
redshift and on the different types of simulations.  In
Sect. \ref{sect:scatter} we quantify the scatter and its nature
(statistical or intrinsic). In Sect.~\ref{sect:bias} we describe the
velocity bias of subhalos and galaxies with respect to the underlying
diffuse component of DM particles.  In Sect.~\ref{sect:Dynamical
  processes in clusters} we look for a signature of the different
dynamical processes which are at work in galaxy systems, on the
velocity distributions of the different tracers of the gravitational
potential.  Finally in Sect.~\ref{sect:disc} we discuss our results
and present our conclusions.

\section{Simulations}
\label{sect:simhalos}

\subsection{Initial conditions}
Our samples of cluster-sized and group-sized halos are obtained from
29 Lagrangian regions, centred around as many massive halos
identified within a large-volume, low-resolution N-body cosmological simulation,
resimulated with higher resolution. We refer to \cite{bonafede2011}
for a more detailed description of the set of initial conditions used
to generate samples of simulated clusters used for our analysis. 

The parent Dark Matter (DM) simulation followed $1024^3$ DM particles
within a box having a comoving side of 1 h$^{-1}$ Gpc, with h the
Hubble constant in units of 100~km~s$^{-1}$~Mpc$^{-1}$.  The
cosmological model assumed is a flat $\Lambda$CDM one, with
$\Omega_{\rm{m}} = 0.24$ for the matter density parameter,
$\Omega_{\rm{bar}} = 0.04$ for the contribution of baryons,
H$_0=$~72~km~s$^{-1}$~Mpc$^{-1}$ for the present-day Hubble constant,
n$_{\rm{s}}=$~0.96 for the primordial spectral index and $\sigma_8 =
0.8$ for the normalisation of the power spectrum.
Within each Lagrangian region we increased the mass resolution and
added the relevant high-frequency modes of the power spectrum,
following the zoomed initial condition (ZIC) technique
\citep{tormen1997}. Outside these regions, particles of mass
increasing with distance from the target halo are used, so that the
computational effort is concentrated on the region of interest, while
a correct description of the large scale tidal field is preserved.
Each high-resolution Lagrangian region is shaped in such a way that no
low-resolution particle contaminates the central ‘zoomed-in’ halo at z
= 0 at least out to 5 virial radii. As a result, each region is
sufficiently large to contain more than one interesting halo with no
contaminants within its virial radius.

Initial conditions have been first generated both for DM-only
simulations.  The mass of DM particles in the zoomed--in regions is
$\rm{m}_{\rm{DM}} = 10^9 \, \rm{h}^{-1} \Msun$. Henceforth we refer to
these simulation as DM-only.  Initial conditions for hydrodynamical
simulations have been generated only in the low--resolution version,
by splitting each particle within the high-resolution region into two,
one representing DM and another representing the gas component, with a
mass ratio such to reproduce the cosmic baryon fraction. The mass of
each DM particle is then m$_{\rm{DM}} = 8.47 \cdot 10^8 \, \rm{h}^{-1}
\Msun$ and the initial mass of each gas particle is m$_{\rm{gas}} =
1.53 \cdot 10^8 \, \rm{h}^{-1} \Msun$.

\subsection{The simulation models}
All the simulations have been carried out with the TreeePM--SPH
{\footnotesize {\sc GADGET-3}} code, a more efficient version of the
previous {\footnotesize {\sc GADGET-2}} code \citep{springel05}. As
for the computation of the gravitational force, the Plummer-equivalent
softening length is fixed to $\epsilon = 5\,\rm{h}^{-1}$ kpc in
physical units below z = 2, while being kept fixed in comoving units
at higher redshift.

Besides the DM--only simulation, we also carried out a set of
non--radiative hydrodynamic simulations (NR hereafter) and two sets of
radiative simulations, based on different models for the release of
energy feedback.

A first set of radiative simulations includes star formation and the
effect of feedback triggered by supernova (SN) explosions (CSF set
hereafter).  Radiative cooling rates are computed by following the
same procedure presented by \cite{wiersma_etal09}. We account for the
presence of the cosmic microwave background (CMB) and for the model of
UV/X--ray background radiation from quasars and galaxies, as computed
by \cite{haardt_madau01}. The contributions to cooling from each one
of eleven elements (H, He, C, N, O, Ne, Mg, Si, S, Ca, Fe) have been
pre--computed using the publicly available {\footnotesize {\sc
    CLOUDY}} photo--ionisation code \citep{ferland_etal98} for an
optically thin gas in (photo--ionisation) equilibrium. Gas particles
above a given threshold density are treated as multiphase, so as to
provide a sub–resolution description of the inter–stellar medium,
according to the model originally described by
\cite{springel_hernquist03}. We also include a description of metal
production from chemical enrichment contributed by SN-II, SN-Ia and
low and intermediate mass stars \citep{tornatore_etal07}. Stars of
different mass, distributed according to a Chabrier IMF
\citep{chabrier03}, release metals over the time-scale determined by
the corresponding mass-dependent life-times (taken from
\citealt{padovani_matteucci93}).  Kinetic feedback contributed by
SN-II is implemented according to the scheme introduced by
\cite{springel_hernquist03}: a multi-phase star particle is assigned a
probability to be uploaded in galactic outflows, which is proportional
to its star formation rate.  In the CSF simulation set we assume
$\rm{v}_{\rm{w}} = 500\vel$ for the wind velocity.

Another set of radiative simulations is carried out by including the
same physical processes as in the CSF case, with a lower wind velocity
of $\rm{v}_{\rm{w}} = 350\vel$, but also including the effect of AGN
feedback (AGN set, hereafter). In the model for AGN feedback, released
energy results from gas accretion onto supermassive black holes
(BH). This model introduces some modifications with respect to that
originally presented by \cite{springel_etal05} (SMH) and will be
described in detail by Dolag et al. (2012, in preparation). BHs are
described as sink particles, which grow their mass by gas accretion
and merging with other BHs. Gas accretion proceeds at a Bondi rate,
while being Eddington--limited. Radiated energy corresponds to a
fraction of the rest-mass energy of the accreted gas. This fraction is
determined by the radiation efficiency parameter $\epsilon_r=0.1$. The
BH mass is correspondingly decreased by this amount. A fraction of
this radiated energy is thermally coupled to the surrounding gas. We
use $\epsilon_f = 0.1$ for this feedback efficiency, which increases
to $\epsilon_f = 0.4$ whenever accretion enters in the quiescent
``radio'' mode and takes place at a rate smaller than one-hundredth of
the Eddington limit \citep[e.g.][]{sijacki_etal07,fabjan_etal10}.

\subsection{The samples of simulated clusters}
\label{sect:samples}
The identification of clusters proceeds by using a catalogue of FoF
groups as a starting point.  The SUBFIND algorithm
\citep{springel01b,dolag2009} is used to identify the main halo, whose
centre corresponds to the position of the most bound DM particle, and
substructures within each FoF group. In the following, we will name
``galaxies'' the bound stellar structures hosted by the subhalos
identified by SUBFIND in the radiative CSF and AGN hydrodynamical
simulations.

In this work we consider all the main halos
with $M_{200}>10^{13} \Msun$ from $z = 0$ to $z = 2$, which contain no
low--resolution particles within $\rm{r}_{200}$. Among these cluster-sized
and group-sized halos, we only retain those with at least five
subhalos more massive than $10^{11} \Msun$ within $\rtwo$.  The number
of selected halos varies at different redshifts and in different
simulation sets, from a minimum of 54 to a maximum of 308.

The subhalos that we consider in our analysis are selected to be more
massive than $10^{11} \Msun$, which corresponds to 72 particles in the
DM simulations. The galaxies we consider in our analysis (in the CSF
and AGN sets) are selected to have a stellar mass $\geq 3 \times 10^9
\Msun$. By choosing this lower limit in stellar mass, we retain all
subhalos more massive than $10^{11} \Msun$ and include many others
with smaller masses. As a consequence, there are more halos with $\geq
5$ galaxies than with $\geq 5$ subhalos (within $\rtwo$). Note that
the effects of the AGN feedback is negligible for galaxies with
stellar masses below the chosen limit since they are generally hosted
within halos where BH particles have never been seeded.


\section{The velocity dispersion - mass relation}
\label{sect:svmrel}

Given a tracer of the gravitational potential of a halo (DM particles,
subhalos, galaxies) it is possible to write a relation between halo
mass and velocity dispersion of the tracer, based on (i) the
definition of circular velocity at $\rtwo$,
$\vtwo=10\;[G\,h(z)\,\Mtwo]^{1/3}$, and (ii) the relation between
$\stwo$ and $\vtwo$. A relation between velocity dispersion and mass
can be derived analytically once the form of the mass density profile
and of the velocity anisotropy profile are given \citep[see e.g. ][
  and the erratum \citealt{mauduit2009erratum}]{mauduit2007}. 

Following \citet{evrard08}, we use the one-dimensional velocity
dispersion $\soneD \equiv \stwo/\sqrt{3}$.  Using the relations
provided by \citet[][eqs.~22--26]{lokas01} we calculate the ratio
between $\soneD$ and $\vtwo$ for NFW mass profiles of concentrations
$c=3$ and 10, and for (constant) velocity anisotropies\footnote{The
  quantities $\sigma_t\equiv[(\sigma_\theta^2 +
    \sigma_\phi^2)/2]^{1/2}$ and $\sigma_r$ are, respectively, the 1D
  tangential and radial component of the 3D velocity dispersion.}
$\beta=1-(\sigma_t/\sigma_r)^2=0$ and 0.5. These values of $c$ and
$\beta$ are typical of group- and cluster-sized halos
\citep[e.g.][]{gao2008,Wojtak2008,Mamon+10}. We find
$\soneD/\vtwo=0.64,0.69$ for $\beta=0$ and $\soneD/\vtwo=0.68,0.70$
for $\beta=0.5$; the larger values are for $c=10$. The average value
we calculate using DM particles for all the halos selected in our
analysis, $\soneD/\vtwo=0.68$, lies within the same range.

Using the range $\soneD/\vtwo=0.64$--0.70 just found, and the
definition of $\vtwo$, we find
\begin{equation}
\label{expected relation fit}
 \frac{\soneD}{\mbox{km s$^{-1}$}} = A_{\rm{1D}} \cdot \left[\frac{\rm{h(z)} \; \Mtwo}{10^{15} \Msun} \right]^\alpha
\end{equation}
with $A_{\rm{1D}} = 1040$--1140 and $\alpha = 1/3$.  Given that the
real, simulated halos are not perfect NFW spheres, and that a halo
concentration depends on its mass and redshift
\citep[e.g.][]{Duffy+08,gao2008}, the real values of $A_{1D}$ and
$\alpha$ can be different, and need to be evaluated from the
data. Moreover, DM particles, subhalos, and galaxies do not
necessarily obey the same $\soneD$--$\Mtwo$ relation. We therefore
evaluate for each cluster of the samples described in
Sect. \ref{sect:samples} the values of $\soneD$ of DM particles,
subhalos, and (for the CSF and AGN simulations) galaxies. The use of
simulations with different baryonic physics implemented allows us to
understand how baryons and different feedback models modify the
scaling relation between velocity dispersion and mass. The $\soneD$
values of DM particles are obtained using the biweight estimator
\citep{beers90}, when at least 15 data points are available.
Otherwise, as suggested by \citet{beers90}, we use the classical
standard deviation. The confidence intervals for the $\soneD$ values
are obtained using eq.~(16) in \cite{beers90}.

\begin{figure*}
  \includegraphics[width=1.6\columnwidth]{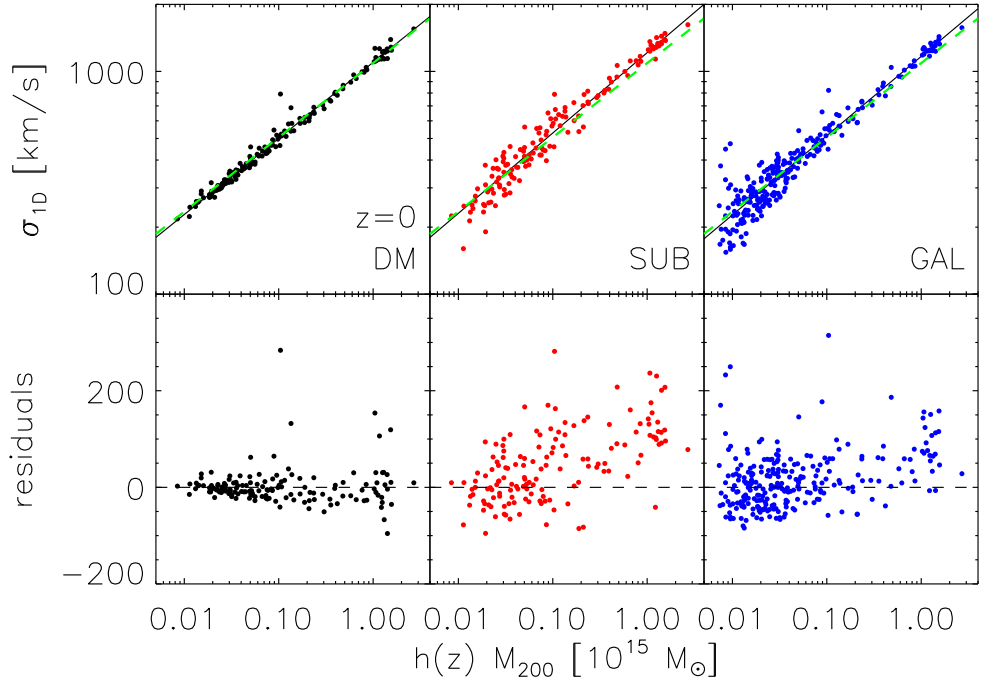}
  \caption{\label{Fig: s vs m} {\em Top panels:} velocity dispersion
    $\soneD$ (km~s$^{-1}$) as a function of halo mass $\rm{h(z)} \,
    \Mtwo$ ($10^{15} \, \Msun$), for DM particles (left panels),
    subhalos (central panels), and galaxies (right panels), at z=0, in
    the AGN simulation sets.  The dashed green line represents the
    theoretically expected relation $\soneD= 1090 \cdot (\rm{h(z)} \,
    \Mtwo)^{1/3}$; the solid line in each panel represents the
    best-fit relation. {\em Bottom panels:} y-axis residuals of the DM
    particles (left), subhalos (centre), and galaxies (right), from
    the DM best-fit relation.}
\end{figure*}

We then perform a linear fit to the $\log(\soneD)$ vs. $\log[\rm{h(z)}
  \, \Mtwo]$ values, with $\soneD$ in units of km~s$^{-1}$ and $\Mtwo$
in units of $10^{15} \, \rm{h}^{-1} \Msun$, for each simulation set,
at several redshifts. The fits were performed with the IDL procedure
\texttt{linfit}, inversely weighting the data by the uncertainties in
the values of $\soneD$. The results of these fits are the values of
the parameters $A_{\rm{1D}}$ and $\alpha$ of eq.~(\ref{expected
  relation fit}).  In Fig.~\ref{Fig: s vs m} and ~\ref{Fig: s vs m
  z126} we show examples of these fits for the AGN simulations at
redshift 0 and 1.26, respectively.

\begin{figure*}
  \includegraphics[width=1.6\columnwidth]{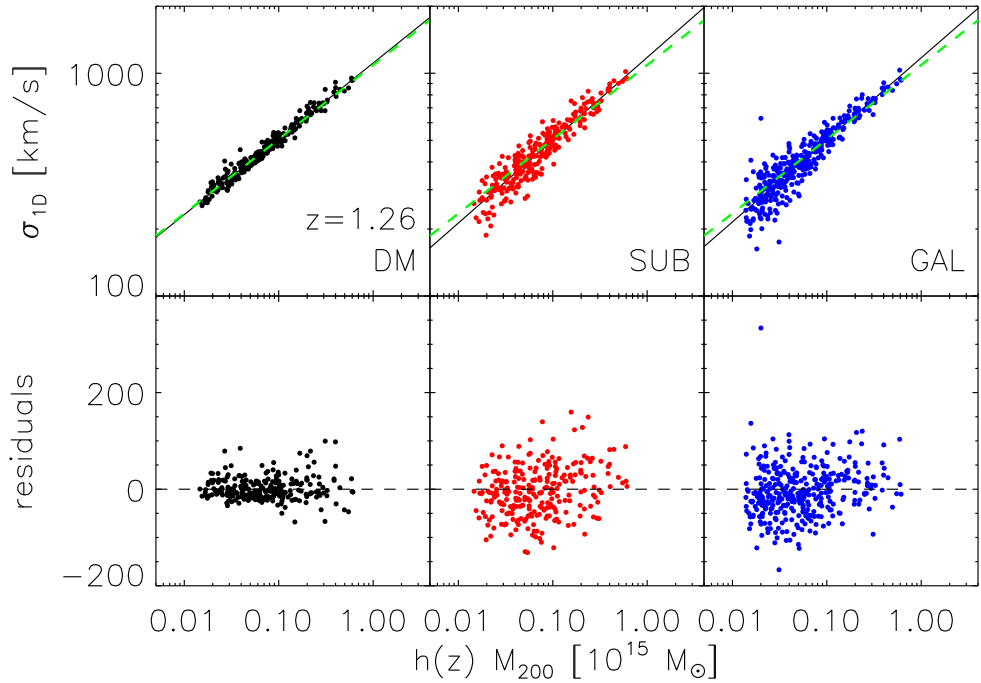}
  \caption{\label{Fig: s vs m z126} Same as Fig.~\ref{Fig: s vs m} but
    for z=1.26.}
\end{figure*}

\begin{figure}
  \includegraphics[width=\columnwidth]{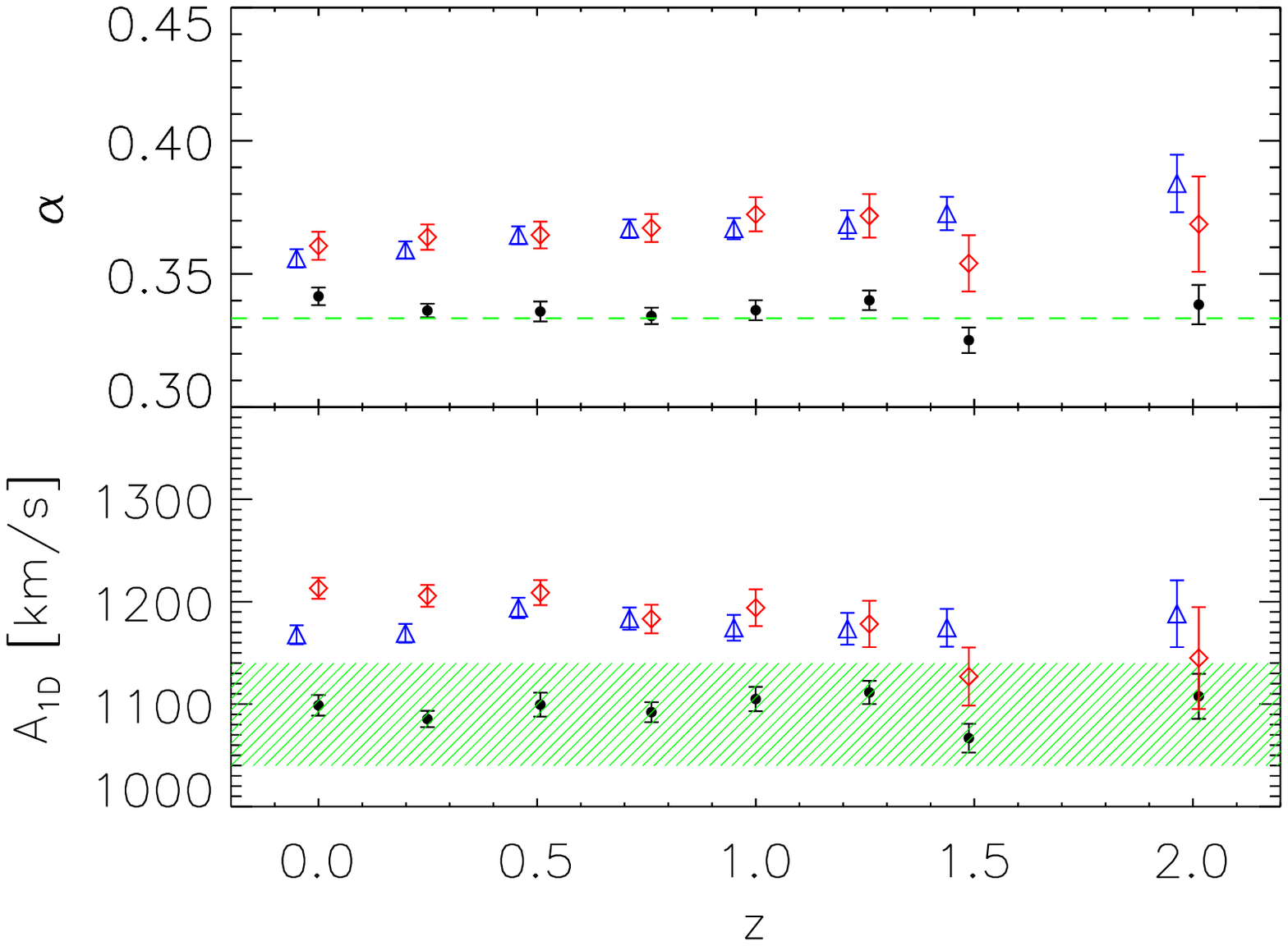}  
  \caption{\label{Fig: fitevolution} Average values of $\alpha$ and
    $A_{\rm{1D}}$ (see eq.~\ref{expected relation fit}) for halos in
    AGN simulations as a function of redshift, for different tracers,
    DM particle (black dots), subhalos (red diamonds), and galaxies
    (blue triangles). The dashed green line in the top panel is the
    theoretically expected value $\alpha = 1/3$, the horizontal shaded
    area in the bottom panel indicates the expected theoretical range
    $A_{\rm{1D}} = 1040$--1140 (see text). The data of galaxies
    have been shifted by $-0.05$ in z for the sake of clarity.}
\end{figure}

\begin{figure}
  \includegraphics[width=\columnwidth]{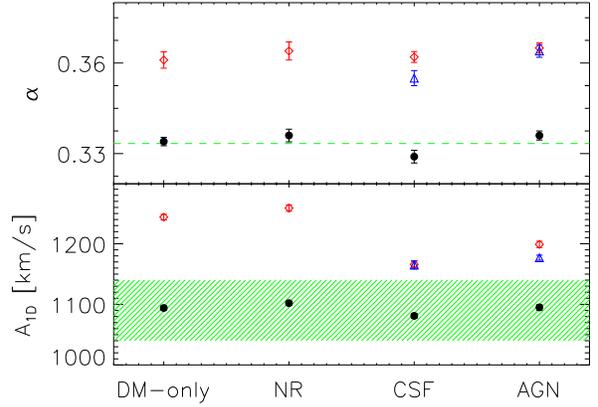}  
  \caption{\label{Fig: comparison} Values of $\alpha$ and
    $A_{\rm{1D}}$ (see eq.~\ref{expected relation fit}) averaged over
    all redshifts for different tracers - DM particle (black dots),
    subhalos (red diamonds), and galaxies (blue triangles) - for the
    different types of simulations. The dashed green line in the top
    panel is the theoretically expected value $\alpha = 1/3$, the
    horizontal shaded area in the bottom panel indicates the expected
    theoretical range $A_{\rm{1D}} = 1040$--1140 (see the text).}
\end{figure}

In Fig.~\ref{Fig: fitevolution} we show the best fitting values of
$A_{\rm{1D}}$ and $\alpha$ as a function of redshift, for different
tracers in the AGN simulation, as an example. The slope $\alpha$ is
confirmed to be consistent with the theoretically expected value
$\alpha = 1/3$. On the other hand, its value is significantly larger
when using subhalos or galaxies as tracers. In any case the values of
$\alpha$ and $A_{1D}$ do not generally show a significant dependence
on z. Only in two cases (subhalos in the DM simulation and galaxies in
the AGN simulation - see Appendix \ref{sect:appendix}) we do find
(marginally) significant correlations between $\alpha$ and z, mostly
driven by the point at z = 2. Even in these cases, $\alpha$ changes
very little with redshift, $\simeq 4$\% for the galaxies in the AGN
simulation from z = 1.5 to z = 0. In fact, a model where $\alpha$ is
constant with z provides an acceptable fit (in a $\chi^2$ sense) to
all cases (also those not shown in Fig.~\ref{Fig: fitevolution}).
Since the variation of $A_{\rm{1D}}$ and $\alpha$ with z is not
significant, we take the weighted averages of their values over all
redshifts to characterise the $\soneD-\Mtwo$ relations of the
different types of simulations and tracers (see Table \ref{Tab: fit
  values} and Fig. \ref{Fig: comparison}).

\begin{table}
\centering
\begin{tabular}{lll}
\hline
 & \multicolumn{1}{c}{$A_{1D}$ (km~s$^{-1}$)} & \multicolumn{1}{c}{$\alpha$}\\ 
\hline 
DM & $     1094\pm       3.7$ & $     0.334\pm    0.0014$ \\
NR & $     1102\pm       3.4$ & $     0.336\pm    0.0021$ \\
CSF & $     1081\pm       4.1$ & $     0.329\pm    0.0021$ \\
AGN & $     1095\pm       4.4$ & $     0.336\pm    0.0015$ \\
\hline
DM sub & $     1244\pm       4.7$ & $     0.361\pm    0.0027$ \\
NR sub & $     1259\pm       5.3$ & $     0.364\pm    0.0030$ \\
CSF sub & $     1166\pm       5.1$ & $     0.362\pm    0.0018$ \\
AGN sub & $     1199\pm       5.2$ & $     0.365\pm    0.0017$ \\
\hline
CSF gal & $     1165\pm       6.7$ & $     0.355\pm    0.0025$ \\
AGN gal & $     1177\pm       4.2$ & $     0.364\pm    0.0021$ \\
\hline \hline 
\end{tabular}
\caption{\label{Tab: fit values} Weighted average values over all
  redshifts of the parameters $A_{1D}$ and $\alpha$ defining the
  $\sigma_{1D} - M_{200}$ relation (see eq. (\ref{expected relation
    fit})) for different simulation sets, for DM particles (top
  table), subhalos (middle table), and galaxies (bottom table).}
\end{table}

In Fig. \ref{Fig: comparison} we show the dependence of the parameters
$\alpha$ and $A_{1D}$ on the physical processes included in the
simulations. When considering DM particles as tracers, the
$A_{\rm{1D}}$ values are well within the theoretically expected range,
and the $\alpha$ values are close to the virial expectation 1/3,
regardless of the baryonic physics implemented in the
simulations. When considering subhalos or galaxies as tracers, the
$\soneD$-$\Mtwo$ relations are significantly steeper ($\alpha > 1/3$)
than for DM particles and than expected theoretically. Furthermore,
while the slope is nearly the same for all the simulation sets, with
$\alpha \simeq 0.36$, the same does not hold for the normalization
$A_{1D}$, this value being smaller when cooling and star formation are
included. In each simulation set, the $\alpha$ and $A_{\rm{1D}}$
values are highest for the subhalos, and lowest for the DM particles,
with the values for the galaxies being closer to those of the
subhalos.

Both the $A_{1D}$ and the $\alpha$ values we find for the DM particles
are in the range of the values listed by \cite{evrard08} and coming
from the analysis of different simulations \citep[see Table 4
  in][]{evrard08}, most of which are DM-only.  The comparison with
\cite{lau2010} is less straightforward, since they did find a
z-dependence of the $\alpha$ parameter for DM particles. Taking an
error-weighted average of the values they found at different
redshifts, we obtain $A_{1D}=1103 \pm 2$ km~s$^{-1}$, $\alpha=0.325
\pm 0.011$ for their non-radiative simulations, and $A_{1D}=1160 \pm
9$ km~s$^{-1}$, $\alpha=0.296 \pm 0.012$ for their radiative
simulation. \cite{lau2010}'s $\soneD$--$\Mtwo$ relations are therefore
flatter than ours, particularly so for the radiative simulations. We
discuss these differences in Sect. \ref{sect:disc}.

\subsection{Scatter}
\label{sect:scatter} 

An analysis of the scatter of the scaling relations is quite important
for cosmological applications. It has been pointed out \citep[see,
  e.g.][]{mortonson2011} that the so called Eddington bias causes the
mass function to shift toward higher values of mass if the scatter in
the scaling relation between mass and mass proxy is not properly taken
into account. Furthermore we want to understand the nature of such
scatter, that is whether it is intrinsic or mainly due to statistical
uncertainties.

\begin{figure*}
  \includegraphics[width=1.6\columnwidth]{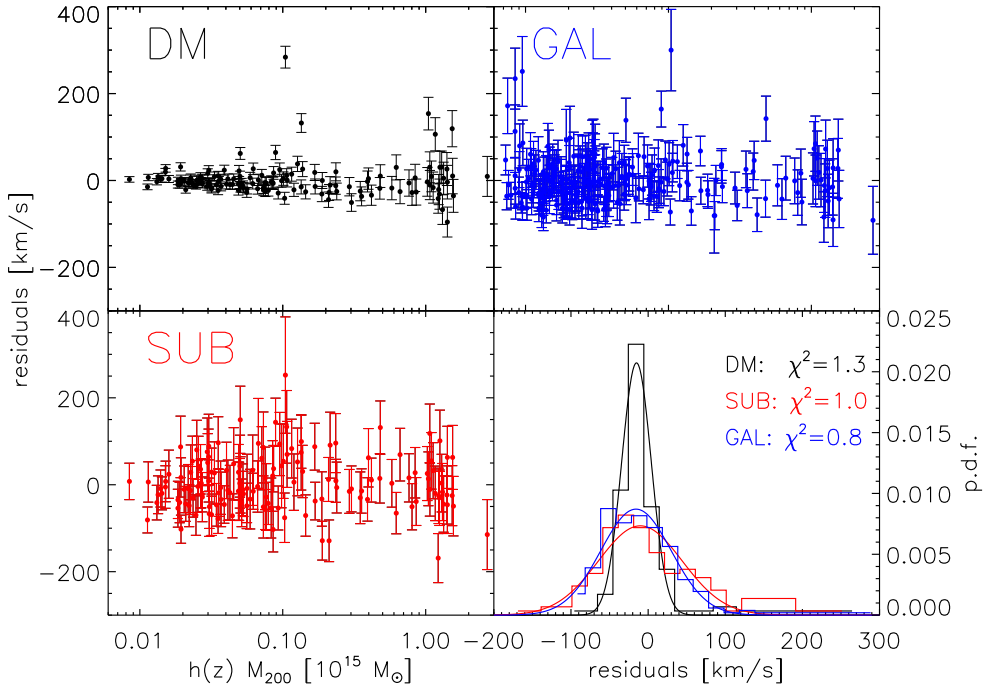}  
  \caption{\label{Fig: scatter} Residuals of the velocity dispersion
    after subtracting the best fit relation eq. (\ref{expected
      relation fit}) for the AGN set at z=0, for the three tracers as
    indicated in the panels. The distributions of residuals for the
    three tracers is also shown in the bottom right panel, along with
    the best fit gaussian curves and the reduced $\chi^2$ of the
    fits.}
\end{figure*}

The best fit relation eq. (\ref{expected relation fit}) has been
subtracted from the values of the velocity dispersion of the clusters
for each simulation at each redshift. The result for the AGN set at
z=0 is shown in Fig. \ref{Fig: scatter} for the three tracers. The
errors are associated to the biweight average value
\citep{beers90}. In this figure the distribution of the residuals is
also shown, along with the best fit gaussian curves and the reduced
$\chi^2$ values of the fits. The residuals appear to be normally
distributed, substructures and galaxies having a wider distribution.

\begin{figure}
  \includegraphics[width=\columnwidth]{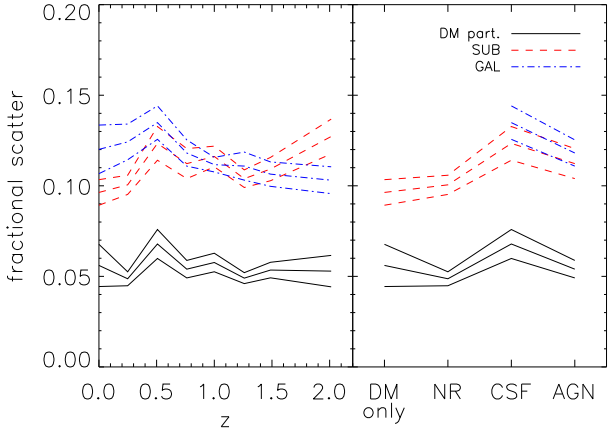}  
  \caption{\label{Fig: scatter vs} \emph{Left panel:} Fractional
    scatter as a function of redshift in the AGN set. Black solid line
    refers to the value computed by using DM particles, red dashed and
    blue dash-dotted by using substructures and galaxies,
    respectively. The thinner lines are the confidence
    intervals. \emph{Right panel:} Fractional scatter as a function of
    the type of simulation at z=0. Lines and colours are the same as
    in the left panel.}
\end{figure}

The scatter appears to be independent of cluster mass, as well as of
redshift and the type of simulation as shown in top panels of
Fig. \ref{Fig: scatter vs}, the only difference being in the tracer
used to build the $\soneD$--$\Mtwo$ relation. When using DM particles
the scatter is around $5\%$, while using substructures or galaxies it
is around $12\%$.

In order to understand the nature of the scatter, that is whether it
is intrinsic or statistical, we have tried to compute the intrinsic
scatter following \cite{williams2010}. Performing a linear fit of the
$\log \sigma_{1D} - \log \Mtwo$ relation, the intrinsic scatter is
considered as a parameter that minimizes $\chi^2=\sum_{i=1}^{N_{clus}}
[y_i-(a \cdot x_i + b)]^2 / [\epsilon_{y,i}^2+a^2 \cdot
  \epsilon_{x,i}^2+\sigma_{int}^2]$, where $y=a \cdot x +b$ is the
linear relation, $\epsilon_{x,i}$ and $\epsilon_{y,i}$ are the
uncertainties on the x and y quantities and $\sigma_{int}$ is the
intrinsic scatter. In order to estimate the value of the intrinsic
scatter and its uncertainty, we performed 1000 bootstrap resamplings
of the couples of values $(\Mtwo,\sigma_{1D})$, each time computing
the intrinsic scatter estimate. In a first time, we have evaluated the
intrinsic scatter in 4 bins of mass, but we found no mass
dependence. Therefore we have evaluated it using all the data. The
values of velocity dispersion evaluated using DM particles have been
obtained using a huge number of objects, hence the statistical
uncertainty is very small. The $5\%$ scatter is therefore entirely
intrinsic. On the other hand velocity dispersions estimates using
substructures and galaxies are based on relatively small numbers of
objects, and statistical uncertainties dominate the scatter. The
resulting intrinsic scatter for these tracers turns out to be quite
small and consistent with the value found for DM particles.

\subsection{Velocity bias}
\label{sect:bias} 
\begin{figure}
  \includegraphics[width=\columnwidth]{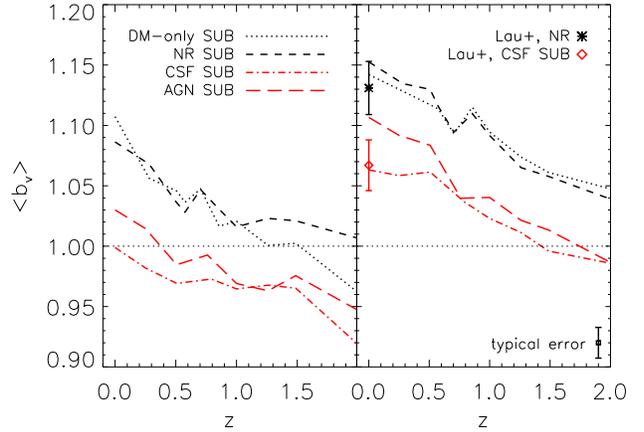}
  \caption{\label{Fig: bias sub} Average values of the velocity bias
    for subhalos, as a function of redshift, for halos in the low-mass
    sample (left panel) and in the high-mass sample (right panel).
    Dotted line refers to substructures in the DM set, short black
    dashed line refers to substructures in the NR set, the other lines
    refer to substructures of the CSF and AGN sets, as indicated in
    the legend.  Error-bars on the bias values are not shown for the
    sake of clarity. A typical error-bar is shown at the bottom-right
    of the right panel. The other points with error bars are the z = 0
    values from \protect\cite{lau2010}. The legend in the right panel
    identifies the different symbols as representative of the
    non-radiative simulations (NR), and the CSF simulations of
    \protect\cite{lau2010}.}
\end{figure}

\begin{figure}
  \includegraphics[width=\columnwidth]{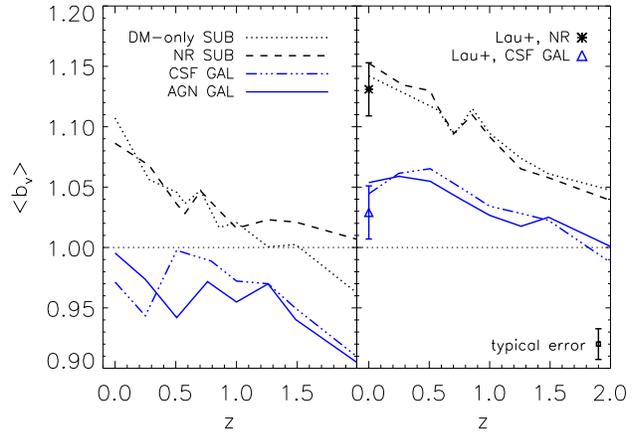}
  \caption{\label{Fig: bias gal} Same as Fig. \ref{Fig: bias sub} but
    for galaxies rather than substructures.}
\end{figure}

The difference between the $\soneD$--$\Mtwo$ relations established for
DM particles, on one side, and subhalos and galaxies, on the other
side implies that subhalos and galaxies have a higher velocity
dispersion than DM particles in high mass halos. Since the relation is
steeper for subhalos and galaxies than for DM particles, the opposite
may occur in halos of low masses.  To examine how $\soneD$ changes in
halos of different masses when using different implementations of
baryonic physics, here we analyse the velocity bias, i.e.  the ratio
between the velocity dispersions of subhalos (galaxies) and DM
particles, $\rm{b_v} = \sigma_{\rm{sub}} / \sigma_{\rm{DM}}$
($\rm{b_v} = \sigma_{\rm{gal}} / \sigma_{\rm{DM}}$, respectively), for
two subsamples of halos, one with masses $\rm{h(z)} \Mtwo < 10^{14}
\Msun$, and the other with masses $\rm{h(z)} \Mtwo > 3 \cdot 10^{14}
\Msun$ (the low- and high-mass samples hereafter). The average
$\rm{b_v}$ values as a function of redshift are shown in
Fig. \ref{Fig: bias sub} and \ref{Fig: bias gal}, for subhalos and
galaxies in different simulation sets, separately for the halos in the
low-mass sample (left panels) and for the halos in the high-mass
sample (right panels).


The bias is on average higher for the high-mass sample halos, and at
lower redshifts. A remarkable change in the bias appears when
introducing cooling and star formation. In fact the biases are higher
for substructures in the DM and in non-radiative simulations (NR) than
in the radiative ones (CSF, AGN), both when using substructures and
galaxies, reflecting again the importance of the cooling and the
feedback processes in the dynamics.

In Fig. \ref{Fig: bias sub} and \ref{Fig: bias gal} we also show the
values found by \cite{lau2010}. At z = 0, the sample of
\citet{lau2010} and ours have similar mass distributions and there is
a good agreement (within the error bars) between their $\rm{b_v}$
values and ours, separately for the different types of simulations and
for the different tracers. The comparison between our data and those
in \cite{lau2010} at $\rm{z} > 0$ is not shown because it is not
straightforward: \cite{lau2010} follow the same halos at different
redshifts by always considering the most massive progenitors of the
halos selected at z = 0, whereas we consider all halos above a given
mass cut at any given redshift.  Taken at face value, the biases they
find at higher z are smaller than those we find, and this difference
might be related to a strong overcooling present in their simulations,
making their subhalos very resistant to tidal disruption.


\subsection{Dynamical processes in halos}
\label{sect:Dynamical processes in clusters}

The above results show how the relative kinetic energy content in
subhalos (or galaxies) and the diffuse DM component varies with
redshift, halo mass, and the type of simulations. Here we provide an
interpretation of these results in terms of the dynamical processes
that are effective in clusters and groups, i.e. dynamical friction,
merging with the central galaxy, tidal disruption, and accretion from
the surrounding large-scale structure, and how these processes depend
on the physics implemented in the simulations.

\begin{figure}
\begin{tabular}{c}
  \includegraphics[width=\columnwidth]{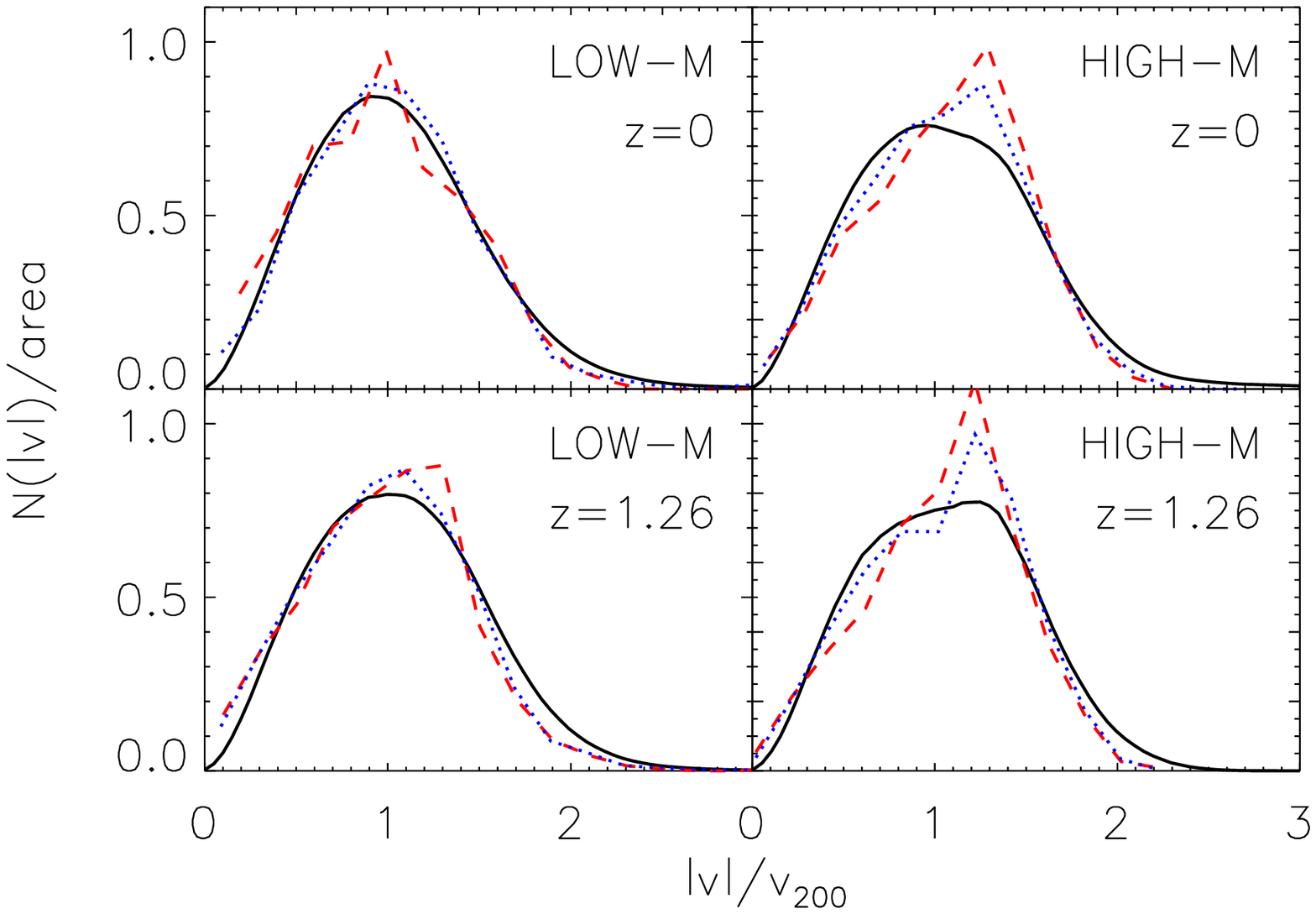}\\
  \includegraphics[width=\columnwidth]{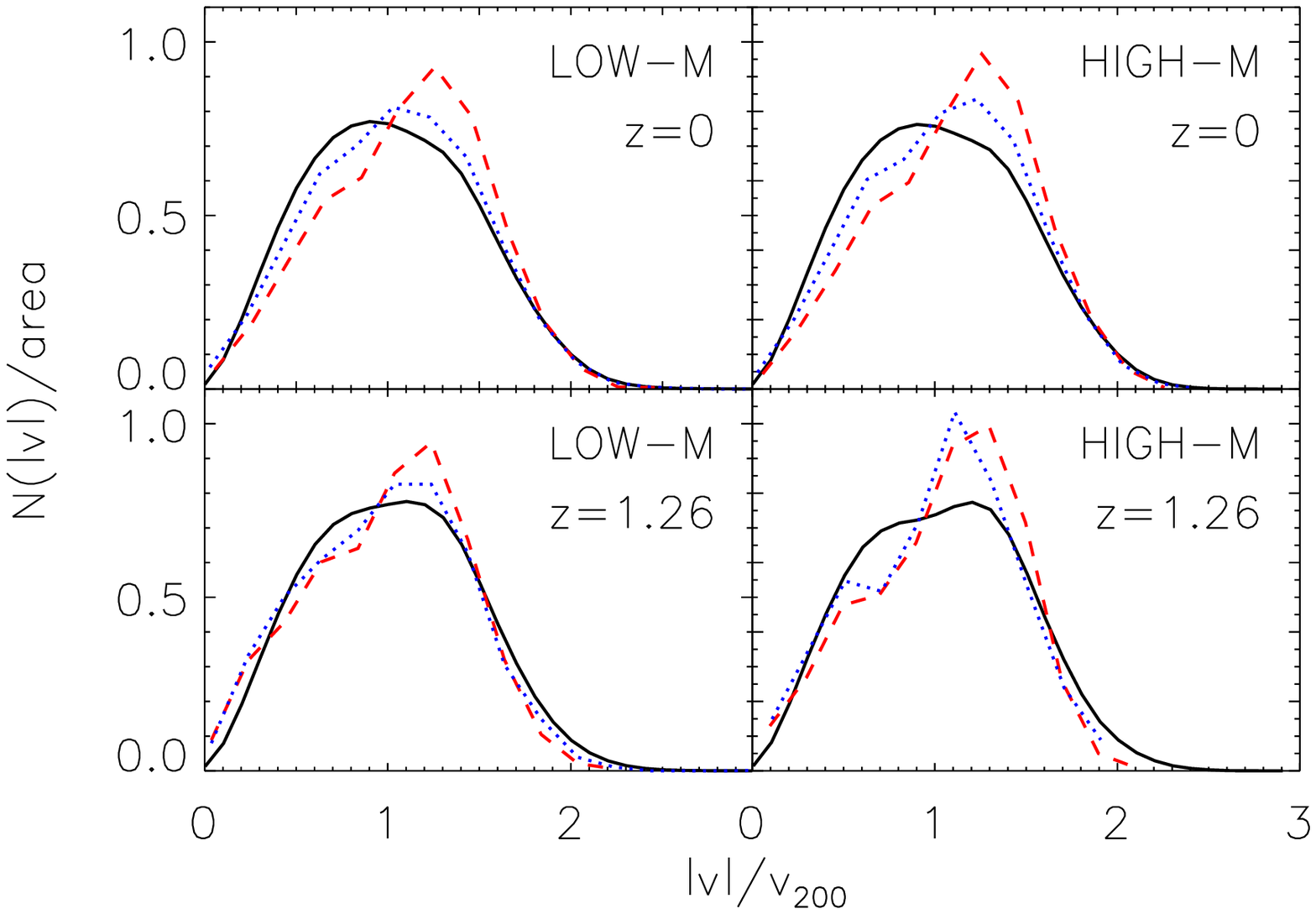}
\end{tabular}
  \caption{\label{Fig: n vs v} \emph{Top panels:} The velocity
    distributions for the CSF simulations for the low-mass sample
    (left panels) and the high-mass sample (right panels) at $z=0$
    (top panels) and $z=1.26$ (bottom panels) for DM particles (solid
    black lines), subhalos (dashed red lines) and galaxies (dotted
    blue lines).  All distributions are normalised to their integrals.
    \emph{Bottom panels: } same as top panels but for the AGN
    simulations.}
\end{figure}

A better understanding of the physical cause of the differences in the
$\soneD$-$\Mtwo$ relation and of the velocity biases analysed in the
two previous sections can be obtained by examining the probability
distribution function for the modulus of the velocity of the different
tracers. These distributions are shown in Fig. \ref{Fig: n vs v} for
the CSF and AGN simulations, at z = 0 and 1.26, separately for the
low- and high-mass samples.

The velocity distribution of DM particles (solid black lines in
Fig. \ref{Fig: n vs v}) appears to be single-peaked, but only for the
low-mass sample. The DM particle velocity distribution for the
high-mass sample is flat-topped and sometimes double-peaked. These
features, present in all the different types of simulations, appear to
be more pronounced for the velocity distributions of subhalos (dashed
red lines) and galaxies (dotted blue lines), and depend on the physics
implemented in the simulations. The difference in the velocity
distributions of DM particles, subhalos, and galaxies is at the origin
of the velocity biases and the differences in their $\soneD$-$\Mtwo$
relations. Hereafter we interpret the effect of the dynamical
processes that shape these velocity distributions.

A flat-topped or double-peaked velocity distribution is the signature
of ongoing mergers and infall of matter into the halos, which tends to
populate the high-velocity part of the velocity distributions
\citep[see, e.g.][]{diemand04,Wang+05,wetzel2011,ribeiro2011}.  With
time, the velocity distribution relaxes via phase-mixing and dynamical
relaxation and approaches a single Maxwellian \citep[e.g.][and
  references therein]{faltenbacher2006,lapi2011}.

The relative importance of the relaxed and unrelaxed velocity
distributions depends on how strong the matter infall rate
is. Higher-mass halos are dynamically younger, and undergo significant
matter infall until more recent times than lower-mass halos
\citep[e.g.][]{lapi2009,faltenbacher2010}. Hence in higher-mass halos
the unrelaxed, high-velocity component is more important than in
lower-mass clusters, as we see in Fig.~\ref{Fig: n vs v}.

The velocity distributions of subhalo and galaxies is different from
those of DM particles, because of additional dynamical processes that
leave the DM particle distributions unaffected. One is the dynamical
friction, leading to a decrease in the orbital energy, hence to an
approach to the cluster centre and a decrease in velocity, eventually
followed by merger with the central cluster galaxy
\citep[e.g.][]{boylan08,wetzel2010}. The other process is tidal
disruption, caused by the integral of tidal interactions leading to
mass losses \citep[e.g.][]{diemand04}. These two processes are
related.  On one hand, dynamical friction becomes ineffective when
tidal mass losses become important \citep[e.g.][]{faltenbacher2007},
because dynamical friction is more effective for more massive
objects. On the other hand, tidal mass losses are enhanced by
dynamical friction, since they are more effective in slow-moving
subhalos (and galaxies) than their fast-moving counterparts
\citep{diemand04}. Hence dynamical friction is likely to be more
effective at the first orbit of a subhalo (or galaxy)
\citep{faltenbacher2007}, while tidal disruption may take several
orbits.

Dynamical friction tends to increase the low-velocity tail of the
distribution. A possible signature of this effect is visible in the
velocity distributions of subhalos and galaxies at high-z in
Fig.~\ref{Fig: n vs v}. On the other hand, the removal of the slowest
subhalos (or galaxies) by tidal disruption or merger with the central
galaxy tends to decrease the low-velocity tail of the distribution.
This is likely to occur in the relaxed component of the global
velocity distribution, since subhalos (and galaxies) in the relaxed
component have spent more time orbiting their parent halos than the
recently infallen, unrelaxed population. The resulting velocity
distribution will then appear to be double peaked, one low-velocity
peak being due to the relaxed component, depopulated by tidal
stripping, and another high-velocity peak due to the infalling
population, that has not yet experienced significant tidal mass
losses.  This is particularly evident in higher-mass halos, in which
the infall rate is higher (at given z) than in lower-mass halos. The
resulting asymmetric velocity distribution is visible in
Fig. \ref{Fig: n vs v} (dashed red and dotted blue lines).

These processes are dependent on the physical processes and feedback
implemented in the simulations. In fact in Fig. \ref{Fig: n vs v} one
can note that the velocity distributions of galaxies and subhalos in
the CSF simulation sets are both closer to the velocity distributions
of DM particles, than the corresponding ones in the AGN simulation
sets. This occurs because baryon cooling tends to make galaxies and
subhalos more resistant against tidal disruption
\citep[e.g.][]{WCDK08,lau2010}. However cooling efficiency is reduced
in simulations including AGN feedback thus making halos less compact
and galaxies more fragile than in the over-cooled CSF simulations.

Yet another difference is visible in Fig. \ref{Fig: n vs v}, namely
the velocity distribution for galaxies in the AGN sets is slightly
closer to the distribution of DM particles than that of subhalos.
This is due to the different mass limit we use to select galaxies and
subhalos (see Sect.~\ref{sect:simhalos}). In our samples, subhalos
selected by means of total bound mass are on average objects of higher
mass than subhalos selected by means of stellar mass, hence they are
subject to stronger dynamical friction, slowing them down and making
them more easily subject to tidal disruption. Our result is
reminiscent of that of \cite{faltenbacher2006} who found that the
velocity distribution of galaxies is quite similar to that of DM
particles when selecting objects by their mass at accretion,
i.e. before tidal stripping operates, as pointed out also by
\cite{lau2010}.  

The processes so far discussed provide a frame for a better
understanding of Fig. \ref{Fig: bias sub} and \ref{Fig: bias gal}. At
fixed galaxy mass, dynamical friction is more efficient in low-mass
clusters and at high redshifts, before tidal stripping decreases
galaxy masses.  Therefore dynamical friction tends to create a
velocity bias $<1$ in low mass, high-z systems, while leaving almost
unaffected high mass clusters.  Tidal stripping is more efficient on
slow moving galaxies, which are stripped and eventually disrupted (or
removed from a mass-limited sample), and operates at all times. This
causes an increase of the bias with time. This has the effect of
erasing the initial dynamical-friction bias in low-mass clusters and
creating a bias $>1$ in high-mass clusters as we approach $\rm{z}=0$.

\begin{figure}
  \includegraphics[width=\columnwidth]{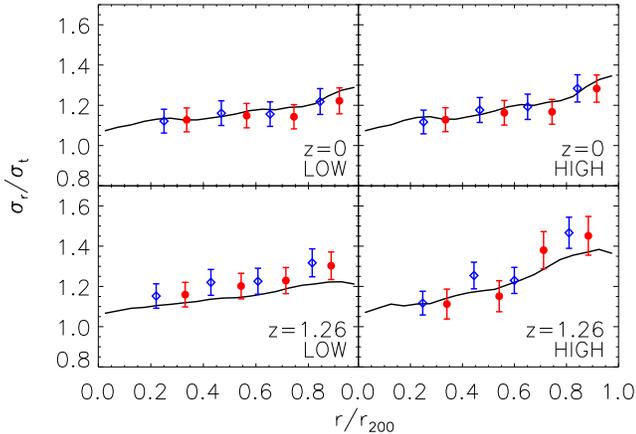}\\
  \caption{\label{Fig: srst vs r} Velocity anisotropy as a function of
    the cluster center, for the AGN set. Top panels refer to z=0,
    bottom panels to z=1.26. Panels on the left are relative to the
    low-mass sample, panels on the right to the high-mass sample. Red
    points refer to the results obtained using subhalos, while blue
    diamonds using galaxies. The four bins are built in such a way
    that within each bin there is the same number of objects. Galaxies
    diamonds have been shifted by -0.05 in redshift for the sake of
    clarity. The black solid line refers to the results obtained using
    DM particles.}
\end{figure}

An important r\^ole in these processes may also be played by galaxy
orbits. In Fig. \ref{Fig: srst vs r} we show the velocity anisotropy
profiles for the different tracers, in low- and high-mass systems, for
z=0 and 1.26. For lack of sufficient number of substructures and
galaxies in each system, the profiles for these tracers are computed
for stacks of all systems, where substructure and galaxy velocities
have been scaled by each system $\rm{v}_{200}$ before stacking. With
this procedure, richer clusters weigh more in the final profile. For
the DM particles we are not limited by poor statistics, so we derive
the anisotropy profiles individually for each halo, and then take an
average. For consistency with what was done for the substructures and
galaxies, the average is weighted by the number of substructures
present in each cluster.  Fig. \ref{Fig: srst vs r} shows that orbits
are more radially anisotropic in high-mass than in low-mass systems
\citep[as already shown by][]{wetzel2011}. Understanding the reason
for this difference (and for the redshift evolution clearly visible in
the same figure) is beyond the scope of this paper, although we
suspect that it might be related to the younger dynamical age of
higher mass clusters \citep[as suggested by, e.g.,][and references
  therein]{biviano2009}. What is relevant in this context is that
galaxies on more radial orbits reach closer to the cluster centre and
therefore suffer from stronger tidal stripping effects. The different
orbital anisotropy of galaxies contribute to create a higher velocity
bias in high-mass relative to low-mass clusters.

\section{Discussion and conclusions}
\label{sect:disc}

We determined the $\soneD$-$\Mtwo$ relation for DM particles,
subhalos, and galaxies in cluster- and group-sized halos extracted
from $\Lambda$CDM cosmological simulations.  We analysed four sets of
simulations carried out for the same halos: one DM-only, one with
non-radiative gas, and another two with gas cooling, star formation
and galactic ejecta triggered by SN winds, one of the two also
including the effect of AGN feedback. 

The main results of our analysis can be summarised as follows.

\begin{itemize}

\item We confirm that the $\sigma_{1D} - M_{200}$ relations for the DM
  particles are consistent with the theoretical expectation from the
  virial relation, assuming NFW halo mass profiles, with reasonable
  values of concentration and velocity anisotropy.  

\item The intercepts at $10^{15} \Msun$ and slopes of the logarithmic
  $\soneD$-$\Mtwo$ relations derived using subhalos and galaxies as
  tracers of the potential are always higher than those derived using
  DM particles. We do not find a significant dependence of the
  $\soneD$-$\Mtwo$ relations on redshift, but we do find a dependence
  on the kind of simulation, the radiative ones having a higher value
  of the normalization.

\item The $\soneD$-$\Mtwo$ relations for the DM particles are
  consistent with those found by \citet{evrard08}. On the other hand
  the relations we find for all the tracers are steeper than those
  derived by \citet{lau2010}.  This difference might be caused by the
  narrower halo mass range explored by \citet{lau2010}.

\item The differences in the $\soneD$-$\Mtwo$ relations for the
  different tracers of the halo gravitational potential and for the
  different physics implemented are due to dynamical processes taking
  place in the halos. In fact dynamical friction and tidal disruption
  act on substructures and galaxies but not on DM particles. Dynamical
  friction slows down a substructure or a galaxy before it suffers
  mass loss due to tidal stripping.  Dynamical friction is therefore
  more efficient at high-z. It is also more efficient in lower-mass
  clusters for given galaxy mass.  Tidal stripping, on the other hand,
  is more efficient in higher-mass clusters, where galaxies move on
  more radial orbits (hence with smaller pericenter radii). As a
  result, velocity biases are created in the substructure and galaxy
  populations, relative to the DM particles, and these biases are
  $\leq 1$ for low-mass systems and $\geq 1$ for high-mass systems,
  and increasing with time, leading to the observed differences in the
  $\soneD$-$\Mtwo$ relations of different tracers.

\item In order to correctly reproduce such processes, a detailed
  implementation of the baryonic physics must be used in the
  simulations. In fact the presence of baryonic matter makes halos
  more resistant against tidal disruption \citep[e.g.][]{WCDK08}. In
  this way in simulations with cooling and star formation there is a
  higher fraction of survivors among the slow moving subhalos,
  reflecting in a lower value of the normalization on the
  $\soneD$-$\Mtwo$ relation. 

\item We analysed the scatter in the $\sigma_{1D}-\Mtwo$ relation,
  finding a value of around $5\%$ for DM particles and $12\%$ for
  substructures and galaxies. The intrinsic scatter, after accountinbg
  for the statistical errors in the $\sigma_{1D}$ measurements,
  appears to be $\lesssim 5\%$, independently of the tracer.

\end{itemize}

Such a small scatter in the $\sigma_{1D}-\Mtwo$ relation potentially
makes $\sigma_{1D}$ a very good proxy for the mass, via inversion of
eq. (\ref{expected relation fit}): $M_{200}/10^{15} \Msun =
(\sigma_{1D}/A_{1D})^{(1/\alpha)}/h(z)$. However, $A_{1D}$ and
$\alpha$ are significantly different for the different tracers (DM
particles, substructures, galaxies). Using the values obtained for one
tracer to infer cluster masses from the $\sigma_{1D}$ of a different
tracer can lead to systematic errors of up to $\sim 30\%$. In
comparison, the effect of using different baryonic physics for the
same tracer has a much smaller effect on the mass estimates obtained
from $\sigma_{1D}$ ($\lesssim 7\%$, see Table \ref{Tab: inverse
  relation}). The presence of scatter, even though small, makes
possible the applicability of the scaling relation only in a
statistical sense, not for mass estimates of single objects.

\begin{table}
\centering
\begin{tabular}{llcc}
simulation & tracer & $\sigma_{1D}=300$ km/s & $\sigma_{1D}=900$ km/s  \\
\hline \hline
DM & DM part &    0.029  &   0.774 \\
NR & DM part &     0.029  &   0.760 \\
CSF & DM part &    0.028  &   0.796 \\
AGN & DM part &  0.029  &   0.775 \\
\hline
DM & sub &    0.027  &   0.567 \\
NR & sub  &  0.027  &   0.552 \\
CSF & sub  &   0.033  &   0.679 \\
AGN & sub   & 0.031  &   0.633 \\
\hline
CSF & gal  &    0.030  &   0.671 \\
AGN & gal  &    0.032  &   0.665 \\
\hline \hline 
\end{tabular}
\caption{\label{Tab: inverse relation} Masses (in $10^{15} \Msun$) of
  clusters at z=0 predicted from the $\sigma_{1D}-\Mtwo$ relation for
  two values of $\sigma_{1D}$, 300 $m/s$ and 900 $m/s$. }
\end{table}

The results of this paper show that good knowledge of the
$\sigma_{1D}-\Mtwo$ relation in 6D phase space is fundamental before
one could apply this relation to observational samples. Projection
effects and the presence of interlopers can significantly affect the
accuracy and reliability of the mass estimate
\citep[e.g.][]{Cen97,biviano06}. We plan to consider these effects in
detail, in a forthcoming work, using simulations with a proper
implementation of baryonic physics and galaxies as tracers.

\section*{Acknowledgements}
We thank the referee, Trevor Ponman, for useful comments and
criticisms that helped us improve the presentation of our results.  We
thank Emanuele Contini and Gary Mamon for useful discussions and
Alvaro Villalobos for the comments on the draft. We are greatly
indebted to Volker Springel for providing us with the non--public
version of {\footnotesize {\sc GADGET-3}}. Simulations have been
carried out in CINECA (Bologna), with CPU time allocated through the
Italian SuperComputing Resource Allocation (ISCRA) and through an
agreement between CINECA and the University of Trieste.  This work has
been partially supported by the European Commission's Framework
Programme 7, through the Marie Curie Initial Training Network
CosmoComp (PITN-GA-2009-238356), by the PRIN-MIUR-2009 grant ``Tracing
the growth of structures in the Universe'', by the PRIN-INAF-2009
Grant ``Toward an Italian network for computational cosmology'' and by
the PD51-INFN grant.  DF acknowledges the support by the European
Union and Ministry of Higher Education, Science and Technology of
Slovenia.

\bibliographystyle{mn2e} 
\bibliography{bibliography}

\clearpage
\appendix
\section{Plots for other models}
\label{sect:appendix}
In this appendix we report the plots showing the redshift dependence
of the slope and the intercept of eq. (\ref{expected relation fit}) as
well as the velocity distributions of the tracers for all simulation
sets not already shown before in the main text.

\begin{figure}
  \includegraphics[width=\columnwidth]{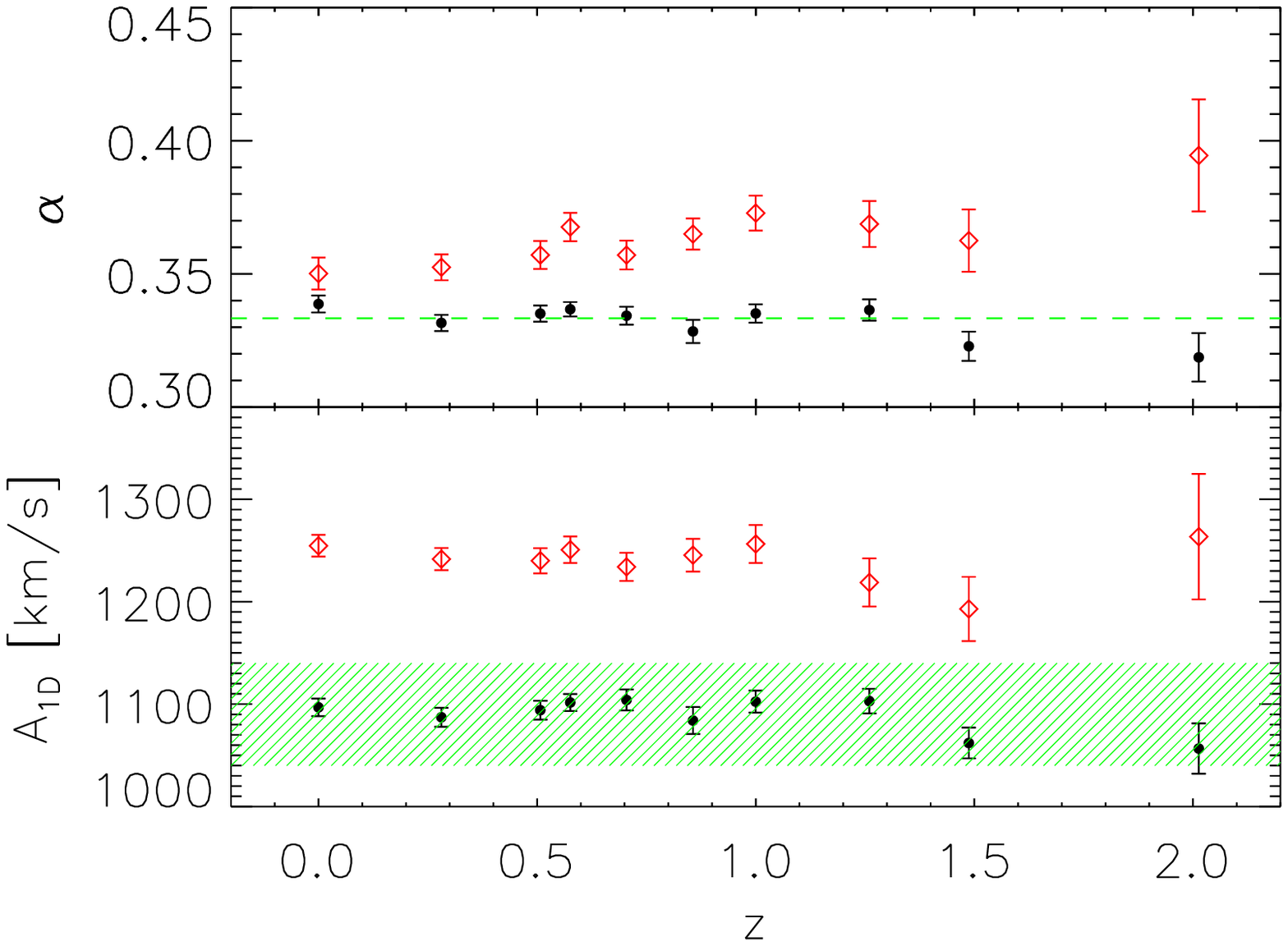}
  \caption{Same as Fig. \ref{Fig: fitevolution} for the DM set.}
\end{figure}

\begin{figure}
  \includegraphics[width=\columnwidth]{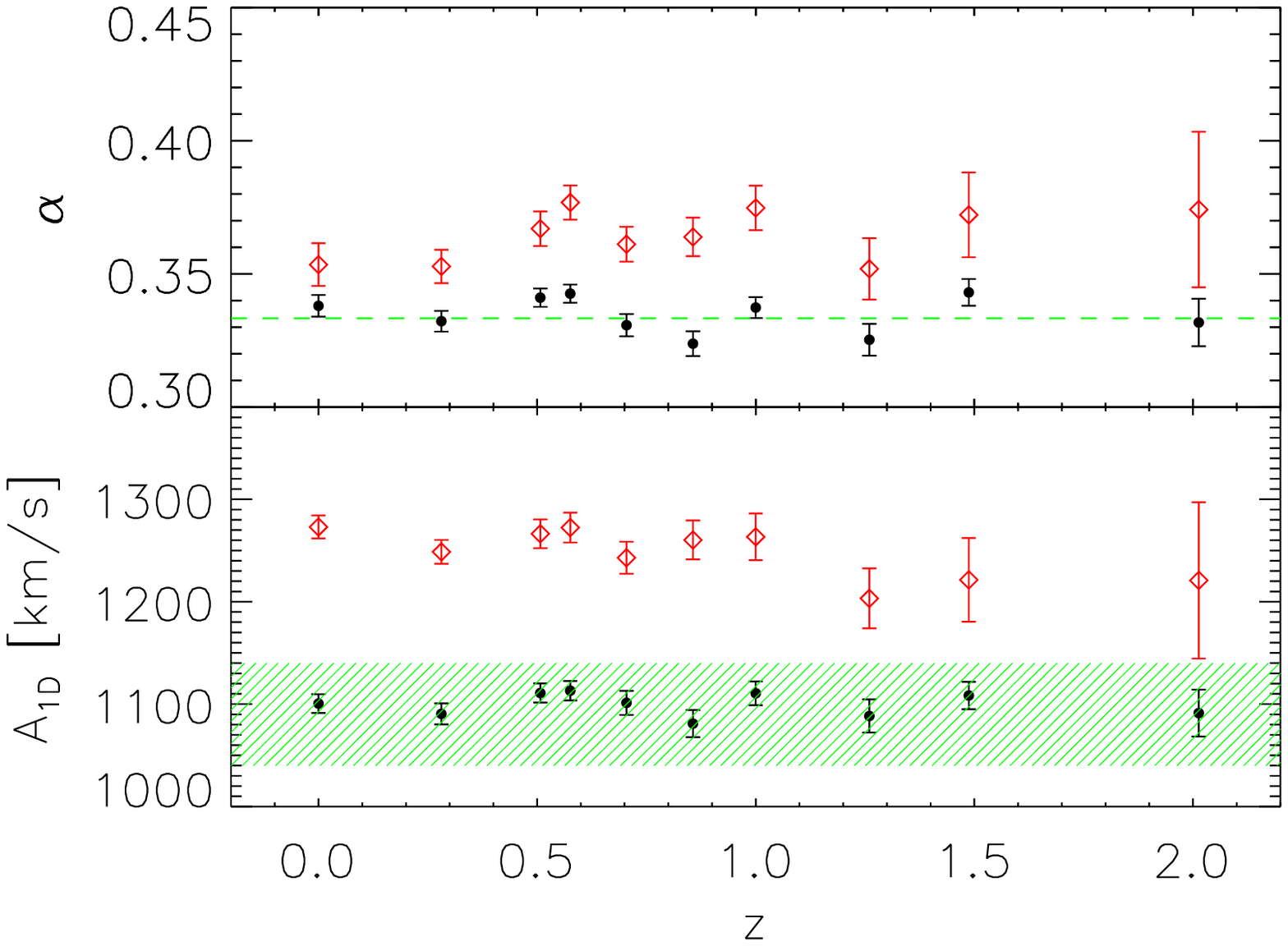}
  \caption{Same as Fig. \ref{Fig: fitevolution} for the NR set.}
\end{figure}

\begin{figure}
  \includegraphics[width=\columnwidth]{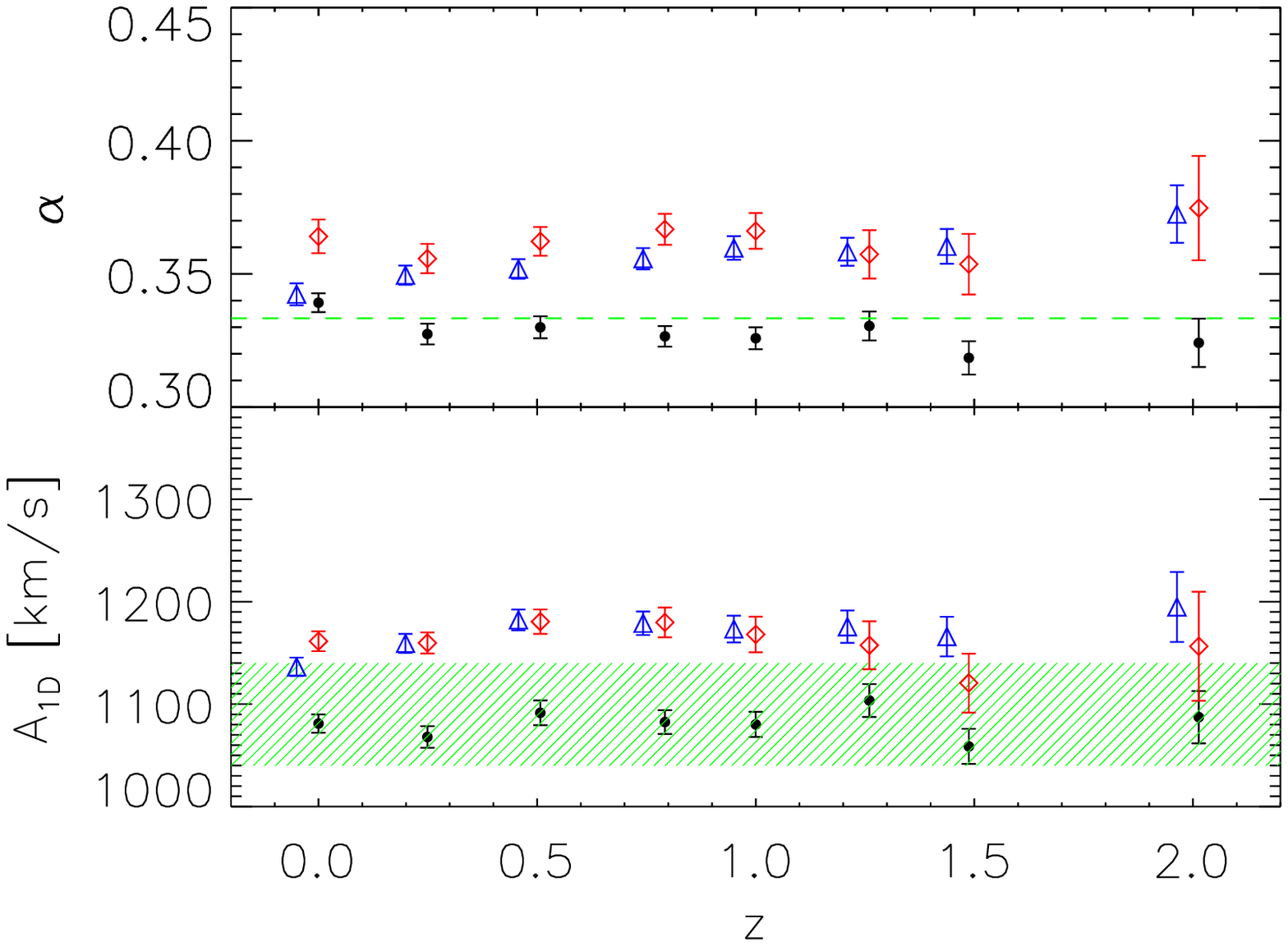}
  \caption{Same as Fig. \ref{Fig: fitevolution} for the CSF set.}
\end{figure}

\begin{figure}
  \includegraphics[width=\columnwidth]{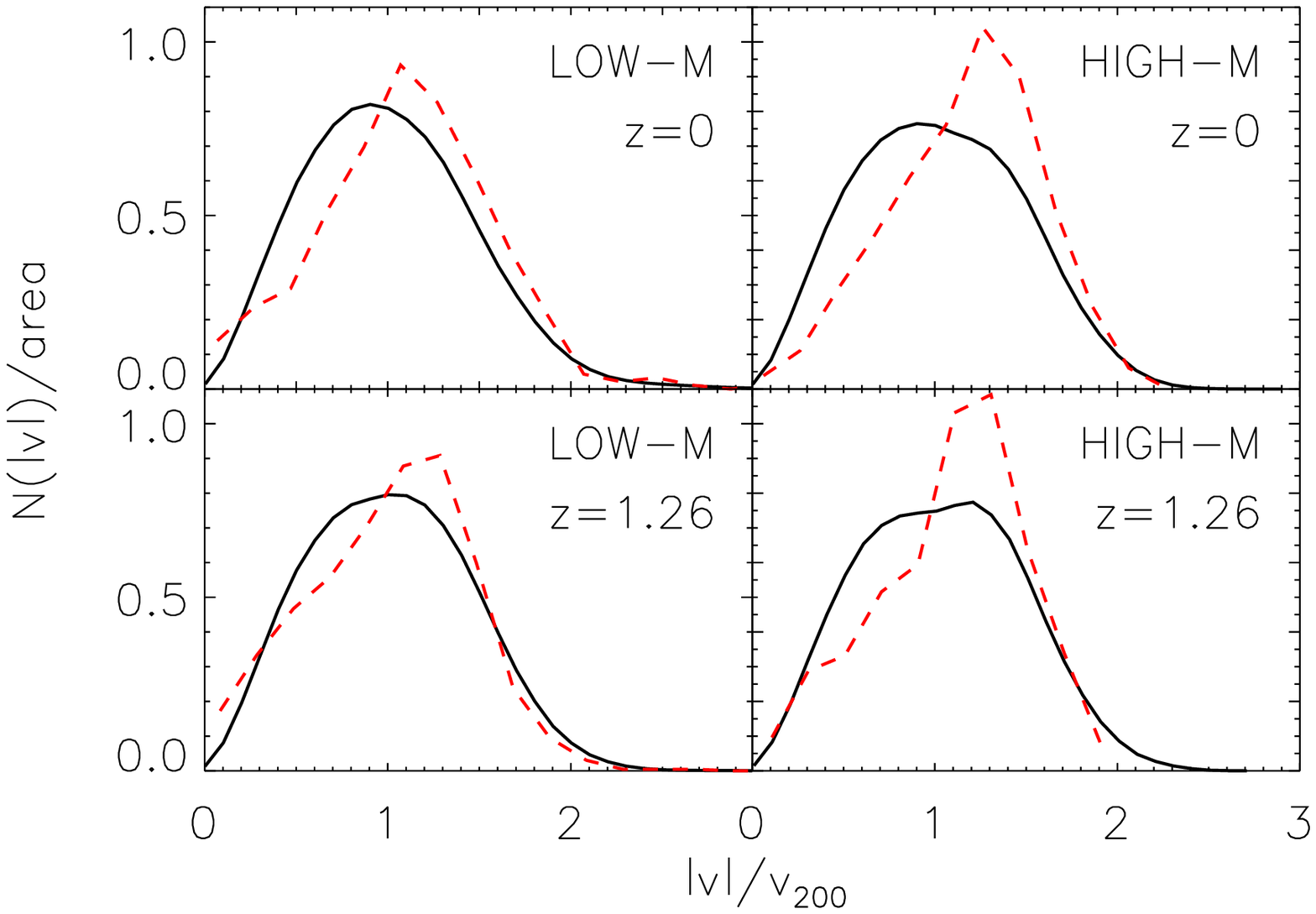}
  \caption{Same as Fig. \ref{Fig: n vs v} for the DM set.}
\end{figure}

\begin{figure}
  \includegraphics[width=\columnwidth]{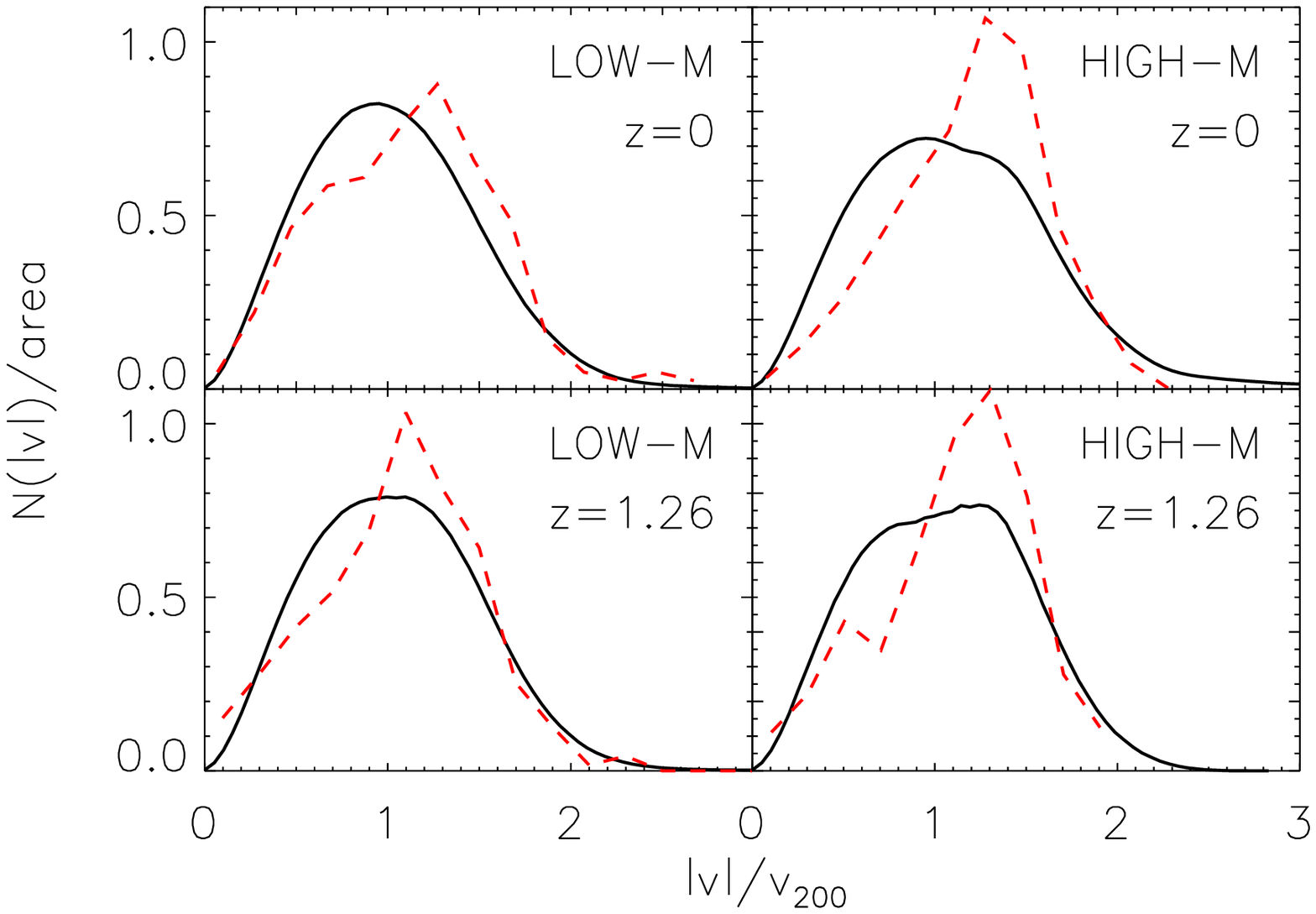}
  \caption{Same as Fig. \ref{Fig: n vs v} for the NR set.}
\end{figure}

\label{lastpage}
\end{document}